\documentclass[12pt]{article}
\usepackage{graphicx}

\input amssym.tex

\begin{document}

\title{Pseudo-Gaussian quantum models}

\author{Ion I. Cot\u aescu\\
{\small \it West University of Timi\c soara,}\\
{\small \it V. P\^ arvan Ave. 4, RO-300223 Timi\c soara, Romania}}

\date{\today}
\maketitle

\begin{abstract}
A new family of one-dimensional quantum models is proposed in terms of new
potentials with a Gaussian asymptotic behavior but approaching to the potential
of the harmonic oscillator when $x\to 0$. It is shown that, in the energy basis
of the harmonic oscillator, the matrix elements of the Hamiltonian operators of
these new models can be derived from generating functionals.

Pacs {03.65.Ge}
\end{abstract}

The progress in algebraic and numerical computation offers one new
possibilities of analyzing new classical or quantum systems that can not be
analytically solved.

The Schr\" odinger equation with an attractive radial Gaussian potential was
analyzed with different perturbation methods leading to numerical results of
convenient accuracy \cite{GG}. Other quantum models with Gaussian barriers were
considered in the theory of electronic transport in mesoscopic systems
\cite{BB,HH}. Some time ago we studied a simple one-dimensional relativistic
model of Gaussian well using generating functionals for finding the matrix
elements of the Klein-Gordon operator in the energy basis of the
non-relativistic harmonic oscillator (HO) \cite{AC}. In this framework
computational methods allowed us to point out the structure of the discrete
energy spectra of the Gaussian wells which indicates that inside the well the
quantum motion is close to an harmonic one.

Here we would like to continue this study in the case of the non-relativistic
quantum system, looking for general one-dimensional models with pseudo-Gaussian
potentials. Our purpose is to define a large family of  models whose potentials
have a Gaussian asymptotic behavior but approaching to the HO potential near
$x\sim 0$. The second objective is to show how  the method of generating
functionals works in this case, helping us to find the matrix elements of the
Hamiltonian operators of our models in the energy basis of HO.

The one-dimensional HO of mass $m$ and frequency $\omega$ has the well-known
Hamiltonian
\begin{equation}
H_0=-\frac{\hbar^2}{2m}\,\frac{d^2}{dx^2}+\frac{m\omega^2 x^2}{2}+V_0
\end{equation}
where $V_0$ is an arbitrary ground energy. The energy is measured in units of
$\epsilon=\hbar\omega$ so that we can  write
\begin{equation}
H_0=\frac{\epsilon}{2}\,N_0\,,\quad N_0=-\frac{d^2}{d\xi^2}+\xi^2+\lambda\,,
\end{equation}
using the dimensionless coordinate  $\xi={x}/{a}$ measured in units of
$a=\hbar/{\sqrt{m\,\epsilon}}$, and denoting $\lambda=2V_0/\epsilon$. The
number operator $N_0$ has the eigenvalues $2n+\lambda+1$ where $n=0,1,2...$ is
the quantum number of the discrete energy levels
$E_n=\epsilon\,[n+\frac{1}{2}(\lambda+1)]$.

\begin{figure}
  \includegraphics[width=13.5cm]{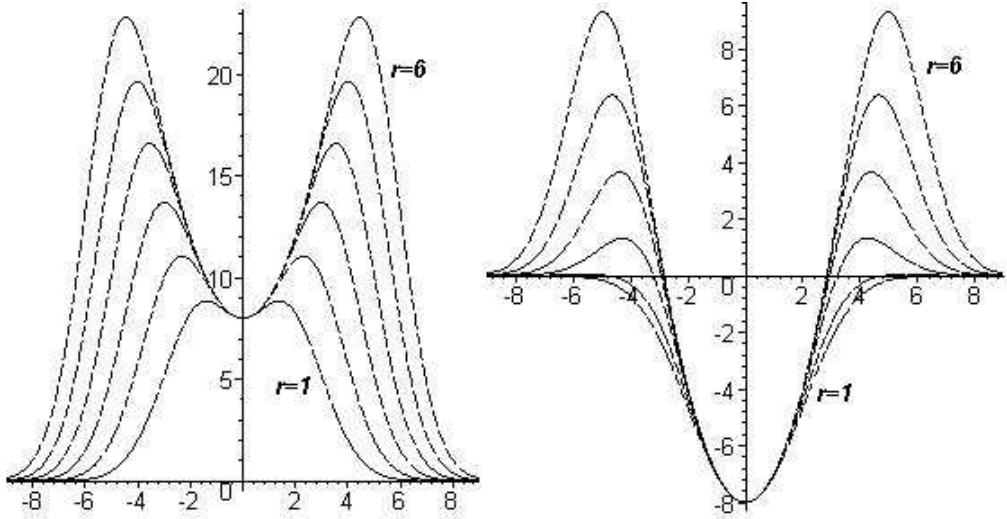}\\
  \caption{The first six pseudo-Gaussian barriers ($\lambda=8$, $\mu=0.2$) and wells
  ($\lambda=-8$, $\mu=0.2$).}\label{f1}
\end{figure}

We propose a Gaussian generalization of HO, defining  pseudo-Gaussian models
(PGM) with new number operators
\begin{equation}
N=-\frac{d^2}{d\xi^2}+ W(\xi)\,,
\end{equation}
in which the dimensionless potentials
\begin{equation}\label{W}
W(\xi)=\left(\lambda+\sum_{k=1}^r C_k \xi^{2k}\right)\exp(-\mu \xi^2)\,,
\end{equation}
are supposed to exhibit Taylor expansions,
\begin{equation}\label{Tay}
W(\xi)=\lambda +\xi^2 + O(\xi^{2r+2})\,,
\end{equation}
but without terms proportional with $\xi^4,\xi^6,...,\xi^{2r}$. It is not
difficult to verify that this condition is accomplished if we take
\begin{equation}\label{Ck}
C_k=\frac{(\lambda\mu+k)\mu^{k-1}}{k!}\,.
\end{equation}
We constructed thus a large family of models, denoted from now by
$(\lambda,\mu)^r$, depending on dimensionless parameters, $\lambda\in {\Bbb R}$
and $\mu>0$, and the integer number $r=1,2,...$  which is called the  {\em
order} of  PGM, unlike the genuine Gaussian potential that is of the order
$r=0$. For $\lambda\ge 0$ the potentials (\ref{W}) are positively defined
representing pseudo-Gaussian barriers  but when $\lambda<0$ we can speak about
pseudo-Gaussian wells of different profiles (Fig. \ref{f1}).

The PGM we consider here have two principal virtues: their potentials (\ref{W})
have a Gaussian asymptotic behavior (provided $\mu>0$) and, in the same time,
the form of these potentials in a neighborhood of $\xi=0$ is similar to that of
the HO potential. Moreover, when the order $r$ increases, the potential
(\ref{W}) closes to the HO potential since, according to Eq. (\ref{Tay}), near
$\xi\sim 0$ these potentials differ among themselves only by terms of the order
$O(\xi^{2r+2})$. An important consequence of this property is that
\begin{equation}
\lim_{r\to \infty}W(\xi)=\lambda +\xi^2\,, \quad \forall\, \xi \in {\Bbb R}\,.
\end{equation}
In other respects, we observe that the PGM becomes HO in the limit of
 $\mu\to 0$ too, when we find again that $W(\xi)\to \lambda+\xi^2$.

The Hamiltonian operators of PGM can be written in terms of physical quantities
using the parameter $\epsilon$ as in the case of $H_0$. Thus we have
\begin{equation}
H=\frac{\epsilon}{2}\,N=-\frac{\hbar^2}{2m}\,\frac{d^2}{dx^2}+V(x)\,,
\end{equation}
where the physical potential,
\begin{equation}\label{pot}
V(x)=\frac{\epsilon}{2}\,W\left(\frac{x}{a} \right)\,,
\end{equation}
depends on the physical coordinate $x$ measured in units of $a$ defined before.
The structure of the energy spectra is determined by the concrete form of the
potential (\ref{pot}). Since all the models $(\lambda,\mu)^r$ have potentials
that satisfy
\begin{equation}
V_{\infty}=\lim_{|x|\to \infty} V(x)=0
\end{equation}
we can conclude that all of them have continuous energy spectra in the domain
$S_c(H)=[0,\infty)$. However, the pseudo-Gaussian wells with $\lambda<0$ give
rise, in addition, to discrete energy spectra in domains $S_d(H)=[V_{min}, 0)$
with $V_{min}=V_0=\frac{1}{2}\,\epsilon\,\lambda<0$. Obviously, the discrete
spectra can have at most a {\em finite} number of energy levels.

In general, the energy eigenfunctions of PGM as well as the discrete energy
levels of the pseudo-Gaussian wells can not be calculated using analytical
methods. Therefore, we must look for appropriate perturbation procedures
leading to efficient symbolic calculations on computers combined with
satisfactory convergent numerical methods.

In the case of our PGM it is convenient to use perturbations in the energy
basis of HO, $\{\left|n\right>\,|\,n=0,1,2...\}$, adopting the technique of
generating functionals related to the generating functions \cite{B}
\begin{equation}
F_{\tau}(\xi)=\left(\frac{1}{\pi}\right)^{1/4}
\exp\left(-\frac{\xi^2}{2}+2\xi\tau-\tau^2\right)
\end{equation}
which yield the HO eigenfunctions normalized in the $\xi$-scale as,
\begin{equation}
u_n(\xi)=\left<\xi|n\right>=\frac{1}{\sqrt{n!\,2^n}}\frac{d^n
F_{\tau}(\xi)}{d\tau^n} _{|\tau=0}\,.
\end{equation}
The matrix elements of an operator $X$ can be derived from the corresponding
generating functional,
\begin{equation}\label{Z0}
Z_{\sigma,\tau}[X]=\int d\xi F_{\sigma}(\xi) [X F_{\tau}](\xi)\,,
\end{equation}
according to the rule
\begin{equation}\label{rule}
\left<m|X|n\right>=\left.\frac{1}{\sqrt{m!\,n!\,2^{m+n}}} {\partial_{\sigma}}^{m}
{\partial_{\tau}}^{n}Z_{\sigma,\tau}[X]\right._{|\sigma=\tau=0}\,.
\end{equation}
In general, the matrix $\left<X\right>$ with the elements (\ref{rule}) is
countable since the energy basis of HO is countable. The advantage of this
method is that for HO or PGM the integral (\ref{Z0}) reduces to some well-known
Gaussian integrals. For example, in the simplest case of HO we obtain the
functional
\begin{equation}
Z_{\sigma,\tau}[N_0]= (1+\lambda+4\sigma\tau) \exp(2\sigma\tau)
\end{equation}
giving rise to the diagonal matrix elements $\left<m|N_0|n\right>=(2n+\lambda
+1)\delta_{nm}$.

The generating functional of an arbitrary model $(\lambda,\mu)^r$,
\begin{equation}
Z_{\sigma,\tau}[N]=Z_{\sigma,\tau}\left[-\textstyle{\frac{d^2}{d\xi^2}}\right]
+Z_{\sigma,\tau}\left[W(\xi)\right]\,,
\end{equation}
 can be also calculated in terms of Gaussian integrals. Starting with two important terms,
\begin{eqnarray}
Z_{ \sigma,\tau}\left[-\textstyle{\frac{d^2}{d\xi^2}} \right]
&=&\left[\textstyle{\frac{1}{2}}-(\sigma-\tau)^2\right]\exp(2\sigma\tau)\,,\\
Z_{\sigma,\tau}\left[\exp(-\mu\xi^2)\right]&=&\frac{1}{\sqrt{\mu+1}}\exp\left[2\sigma\tau
-\frac{\mu}{\mu+1}(\sigma+\tau)^2\right]\,,
\end{eqnarray}
and the identity $\xi^{2k}\exp(-\mu\xi^2)=(-1)^k\partial_{\mu}^k\,
\exp(-\mu\xi^2)$,  we obtain the final result
\begin{equation}\label{Z}
Z_{\sigma,\tau}[N]=\tilde Z_{ \sigma,\tau}[N]
\exp(2\sigma\tau)\,.
\end{equation}
where
\begin{eqnarray}\label{ZZ}
\tilde Z_{\sigma,\tau}[N]&=& \frac{1}{2}-(\sigma-\tau)^2 \nonumber\\
&+& \left[\lambda +\sum_{k=1}^r(-1)^k C_k
\partial_{\mu}^k \right]\left\{\frac{1}{\sqrt{\mu+1}}\exp\left[
-\frac{\mu}{\mu+1}(\sigma+\tau)^2\right]\right\}\,,
\end{eqnarray}
depends on the coefficients (\ref{Ck}). Now a nice exercise is to show directly
that
\begin{equation}
\lim_{r\to \infty}Z_{\sigma,\tau}[N]=\lim_{\mu\to
0}Z_{\sigma,\tau}[N]=Z_{\sigma,\tau}[N_0]\,.
\end{equation}

The method of generating functionals offers one the opportunity of using
computers for finding the matrix elements of the operators $N$ in the energy
basis of HO. The first step is to derive the generating functional (\ref{Z})
for a PGM of a given order $r$ calculating the function (\ref{ZZ}) under Maple
or Mathematica. Furthermore, one must choose a concrete model
$(\lambda,\mu)^r$, with fixed parameters, for which one has to calculate on
computer the matrix elements of $N$ according to the general rule (\ref{rule}).
Thus one obtains a finite-dimensional block $\tilde{\left<N\right>}$
representing a truncation of the matrix $\left<N\right>$ which is countable.
Finally, this block can be used for extracting physical results (e. g. the
energy levels in pseudo-Gaussian wells). We note that in this approach the
truncation of the matrix $\left<N\right>$ takes over the role of the standard
perturbation procedure. Therefore, increasing the dimension of the block
$\tilde{\left<N\right>}$ one may improve the accuracy of results.

Our preliminary tests on computers show that the numerical results obtained for
our PGM are satisfactory convergent. Moreover, interesting behaviors can be
observed even in the case of the models of lowest orders. Thus, it seems that
the first discrete energy levels of the pseudo-Gaussian wells tend to respect
the rules of HO while at the limit of separation between the discrete and
continuous spectra we meet again the specific effect reported in \cite{AC},
namely an inflexion point of the function giving the dependence of the
eigenvalues of $\tilde{\left<N\right>}$ on the number of level.

Finally, we conclude that the potentials proposed here realize a graduate
balance (depending on $r$) between the short-range HO dynamics  and the
Gaussian asymptotic behavior. We hope that these models should be useful for
designing new quantum electronic devices in semiconductor heterostructures.

%\newpage

\subsection*{Acknowledgments}

We are grateful to Erhardt Papp for interesting and useful discussions on
closely related subjects.

\end{document}